\documentclass[prb,twocolumn,,amsmath,amssymb,floatfix]{revtex4}
\usepackage{graphicx}
\usepackage{bm}

\newcommand{\be}{\begin{equation}}
\newcommand{\ee}{\end{equation}}

\def \be{\begin{equation}}
\def \ee{\end{equation}}
\def \ba{\begin{array}}
\def \ea{\end{array}}
\def \beq{\begin{eqnarray}}
\def \eeq{\end{eqnarray}}

\def \a{{\alpha}}
\def \t{{\theta}}

\def \d{{\delta}}

\def \av#1{{\langle#1\rangle}}

\begin{document}

\title{Full quantum distribution of contrast in interference experiments between
interacting one dimensional Bose liquids}

\author{Vladimir Gritsev$^1$, Ehud Altman$^{2}$, Eugene Demler$^1$, Anatoli Polkovnikov$^3$}
\affiliation {$^1$Department of Physics, Harvard University, Cambridge, MA 02138\\
$^2$ Department of Condensed Matter Physics, The Weizmann Institute
of Science Rehovot, 76100, Israel\\
$^3$ Department of Physics, Boston University, Boston, MA 02215}

\begin{abstract}
We analyze interference experiments for a pair of independent one
dimensional condensates of interacting bosonic atoms at zero
temperature.  We show that the distribution function of fringe
amplitudes contains non-trivial information about non-local
correlations within individual condensates and can be calculated
explicitly using methods of conformal field theory. We point out
interesting relations between these distribution functions, the
partition function for a quantum impurity in a one-dimensional
Luttinger liquid, and transfer matrices of conformal field
theories. We demonstrate the connection between interference
experiments in cold atoms and a variety of statistical models ranging
from stochastic growth models to two dimensional quantum gravity. Such
connection can be used to design a quantum simulator of unusual
two-dimensional models described by nonunitary conformal field
theories with negative central charges.
\end{abstract}

\date{\today}

\maketitle

\section{Introduction}

The hallmark of a Bose-Einstein condensate (BEC) is the existence of a
well defined macroscopic phase. Indeed experiments with
large condensates display robust matter wave interference with
negligible fluctuations in the fringe
contrast~\cite{ketterle}. From this point of view, large
three dimensional condensates may be thought of as classical
objects. However, there is a continuous range of possibilities
intermediate between perfect condensates on one side and systems that
do not display an interference pattern, like high temperature thermal
gases, on the other~\cite{pad}. For example, one-dimensional
interacting bosons display interference patterns with a reduced
contrast and with non negligible shot to shot fluctuations
of the fringe contrast~\cite{schmiedmayer}.
In this paper we analyze the full distribution
associated with this quantum noise, and discuss what it can tell us
about the underlying strongly correlated state.

By virtue of its direct connection to the concepts of quantum
measurement, study of quantum noise has deepened our understanding in
a variety of areas. Understanding the noise in photo
detection~\cite{photon} prompted the creation of nonclassical states
of light and led to the development of quantum optics. In mesoscopic
electron systems, current fluctuations contain information that is not
available in simple transport measurements. For example, they can be
used to distinguish electrical resistance due to diffusive scattering from
the resistance due to point contact tunneling (see
Ref.~[\onlinecite{meso}] for a review).
Single atom detectors have been used recently to perform Hanburry-Brown-Twiss
experiments with cold atoms \cite{Esslinger,Aspect}.
Finally, analysis of noise correlations
in time of flight experiments of ultra cold atoms has been
proposed~\cite{pra} and tested~\cite{greiner,bloch}, promising a
powerful new technique to access many-body correlations in such
systems. In these applications, the quantities of direct interest
are contained in the first few moments of the noise distribution,
such as the noise power spectrum. However deeper insights into
quantum systems may be gained by obtaining the full statistics of the
fluctuations.

One of the central problems in the field of ultra cold atoms is
finding new ways of characterizing many-body quantum states.  In
this paper we demonstrate that analysis of the distribution function
of contrast in interference experiments between interacting one
dimensional Bose liquids provides a novel probe of non-local
correlations and entanglement present in these systems. It is known
that the average amplitude of interference fringes can be used to
extract two point correlation functions in fluctuating condensates
\cite{pad}. This idea has been successfully employed by Hadzibabic
et al. to measure the Kosterlitz-Thouless transition in two
dimensional systems \cite{hadzibabic}. Interference experiments,
however, contain more information than the average value of the
contrast. Each observation of the interference pattern is a
classical measurement of a quantum mechanical state, so the result
of each individual measurement will be different from the average
value. As we discuss below, higher moments of the distribution
function of interference amplitudes correspond to high order
correlation functions. Hence the knowledge of the entire
distribution function reveals global properties of the system that
depend on non-local correlation functions of arbitrarily high order.
So far, the discussion of full counting statistics has been limited
mainly to systems of non interacting particles~\cite{levitov}. The
main result of this paper, by contrast, is an expression for the
full distribution of amplitudes of interference fringes arising from
one dimensional {\em interacting} Bose liquids, that can be
described by a Luttinger liquid.

The approach used in this paper for analyzing interference
experiments can be generalized to a variety of other measurements in
cold atom systems. For example, one can analyze fluctuations in
particle number in systems with pairing~\cite{belzig} and
fluctuations in magnetization in Mott states of atoms with magnetic
exchange interactions. In both cases distribution functions will
contain non-trivial information about underlying many-body states.
One can also consider time dependent phenomena such as evolution of
phase coherence between a pair of condensates coupled by a finite
tunneling amplitude~\cite{schmiedmayer}.  In this paper we focus on
the distribution function of contrast in interference experiments
between independent Bose liquids in their ground states. Already in
this case we find a non-trivial evolution of the distribution
function from the non-fluctuating perfect contrast for the case of
non-interacting atoms to the Poissonian distribution of contrast
for atoms in the regime of infinitely strong repulsion. We think
that our work constitutes one of the first steps in the direction of
developing a new method for characterizing interacting many-body
quantum states of cold atoms using full distribution function of
some extensive quantum operator, such as the total
magnetization or the interference amplitude operator
defined in equation (\ref{Ap2}).  Different kinds of cold atom systems and
experimental probes can be studied following this general approach.

We start our analysis by establishing the relation between the
probability distribution function of the interference amplitude and
the partition function of a boundary sine-Gordon model. Using
methods of conformal field theory, we reduce this problem to that of
finding a spectral determinant of a simple one-dimensional,
single-particle Schr\"{o}dinger equation. We solve this problem
numerically, as well as analytically using the WKB approximation, to
obtain the desired fringe distribution for any value of the
Luttinger parameter.
\begin{figure}[ht]
\includegraphics[width=8.5cm]{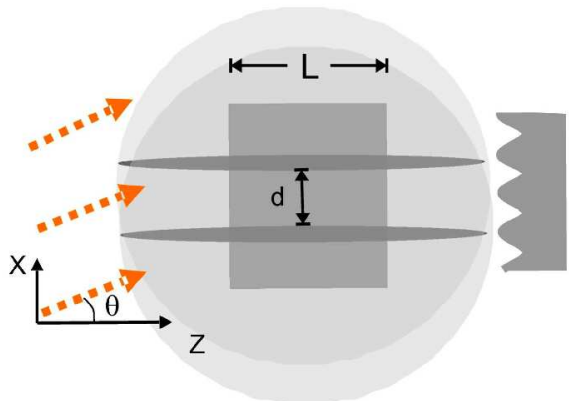}
\caption{Schematic view of the possible experimental setup, which
produces interference pattern between two independent 1D
condensates. } \label{fig:tubes}
\end{figure}

It is interesting to point out that one can take the alternative
point of view on the relation between interference experiments with
BEC and the quantum impurity problem. By measuring the distribution
function of the interference amplitudes experimentally one obtains
the full partition function of the boundary Sine-Gordon problem (see
eq. (\ref{z1}) below). This model describes a range of interesting
problems such as an impurity in a Luttinger
liquid~\cite{kanefisher}, the asymmetric Kondo model~\cite{FLS2},
and the tunneling of a particle in the presence of dissipation
within a Caldeira-Leggett approach~\cite{caldeira-leggett}. As we
discuss below, interference experiments can be used to obtain the
partition function of these systems not only in the ground state but
also in the non-equilibrium regimes (e.g. in the presence of a
finite voltage). Hence interference experiments can be considered as
a quantum solver of these non-trivial many-body problems.

Another surprising implication of our analysis is that interference
experiments with one dimensional cold atoms can be used as a quantum
simulator of several fundamental problems in physics.
This connection relies on the fact that many models of systems with
critical behavior in two-dimensional statistical mechanics, one
dimensional field theories and many-body quantum systems can be
described by continuum theories with conformal invariance.  Basic
ingredients of any conformal field theory are the central charge and
conformal dimensions of operators. One dimensional quantum systems
of interacting particles have positive central charges and
correspond to the so-called unitary class. On the other hand, there
is a non-unitary class of models which have negative central
charges. Such models appear in contexts as different as two
dimensional quantum gravity and stochastic growth models and they
have very unusual properties (see e.g.
Ref.~[\onlinecite{difrancesco}]). In this paper we demonstrate that
interference experiments with one dimensional condensates can be
used to analyze models in nonunitary universality class by virtue of
the {\it exact} mathematical relation between the distribution
function of interference amplitude and the so-called Q-operators of
conformal field theories.  In Fig.~\ref{c_k} below we show a diagram
which illustrates several models that can be studied using
interference experiments.

\section{Full Quantum Distribution of the Interference Contrast}

The setup for the interference experiments we consider is sketched
in Fig.~\ref{fig:tubes}. Two independent quasi-condensates are
allowed to expand in the transverse direction. After sufficient time
of expansion, the integrated density profile is measured by the
imaging beam which is sent at an angle $\t$ to the condensate axis.
This setup is quite common for the interference experiments in cold
atoms and was already realized by several
groups~[\onlinecite{hadzibabic,schumm}].

Everywhere in this paper we consider the two condensates to be
identical, although our analysis can be generalized to the case of
different condensates (see also Ref.~[\onlinecite{niu}]). We assume
that before the expansion, atoms are confined to the lowest
transverse channels of their respective traps and that the optical
imaging length $L$ (which is smaller than or equal to the size of
the system in the axial direction) is much larger than the coherence
length of the condensates. This allows us to use an effective
Luttinger liquid description of the interacting
bosons~\cite{cazalilla}.  The operator corresponding to the
interference signal of the two condensates~\cite{pad} is given by
\be
A_{\tilde p}=\!\!\int_0^L\!\!dz \, a_1^\dagger(z) \, a_2(z)
\,\mathrm e^{i\tilde{p}z}.
\label{Ap2}
\ee
Here $a_1$ and $a_2$ are the bosonic operators in the two systems
before the expansion, and the integrals are taken along the
condensates. The seeming winding of the relative phase between the
two systems, described by the exponential term in equation
(\ref{Ap2}), can either come from the measurement process itself or
from the actual motion of the condensates~\cite{pad}. If the
condensates are at rest, we have $\tilde p=\left(m d/\hbar
t\right)\tan\t$, where $m$ is the atom's mass, $d$ is the separation
between the two condensates and $t$ is the time when the measurement
was done after the free expansion started. With some abuse of
terminology we will call $\tilde{p}$ the relative momentum. When
condensates one and two are independent, one finds that the
expectation value of $\langle A_{\tilde p} \rangle$ vanishes,
however $\langle |A_{\tilde p}|^2 \rangle$ is finite. This means
that individual measurements show a finite amplitude of interference
fringes, however their phase is completely unpredictable. Higher
moments of the interference fringe amplitude are given by~\cite{pad}
\begin{eqnarray}
\langle |A_{\tilde p}|^{2n} \rangle = A_0^{2n} Z_{2n}^{(p)},\;
\mbox{where}\;A_0=\sqrt{C\rho\,\xi_h^{1/K} L^{2-1/K}},
\label{HigherMoments}
\end{eqnarray}
where $C$ is a constant of order unity, $\rho$ is the particle
density in each condensate, $\xi_h$ is the short range cutoff equal
to the healing length, and $K$ is the Luttinger parameter describing
the interaction strength. In this letter we assume that $K>1$, which
is always the case for  bosons with $\delta$-function type repulsive
interactions~\cite{note1}. Coefficients $Z_{2n}$ in
Eq.~(\ref{HigherMoments}) are given by\cite{note}:
\beq
&&Z_{2n}^{(p)}(K)=\int_0^{2\pi}\!\dots\!\int_0^{2\pi}\prod_{i=1}^{n}{du_i\over
2\pi}{dv_i\over 2\pi}e^{i2p\sum_{i}(u_{i}-v_{i})}\nonumber\\
& &\left|{\prod_{i<j} 2\sin\left({u_i-u_j\over 2}\right)\prod_{k<l}
2\sin\left({v_k -v_l\over 2}\right)\over
\prod_{i,k}2\sin\left({u_i-v_k\over 2}\right)}\right|^{1/K},
\label{z2n}
\eeq
where $p$ is the relative momentum measured in units of
$2\pi/L$: $p=\tilde p L/2\pi$.

Coefficients $Z_{2n}^{(p)}$ originally appeared in the grand canonical
partition function of a neutral two-component Coulomb gas on a
circle
\be
Z_{p}(K,x)=\sum_{n=0}^{\infty} {x^{2n}\over (n!)^2} Z_{2n}^{(p)}(K).
\label{z}
\ee
Here $x$ is the fugacity of Coulomb charges and $Z_{2n}$ describe
contributions from configurations with $2n$ charges (i.e.  canonical
partition functions). The partition function (\ref{z}) with $K>1$
and $p$ being half-integer describes several problems in statistical
physics (see Ref.~[\onlinecite{FLS1}] and references therein). In
particular, it describes an impurity in a one-dimensional
interacting electron liquid. At low energies this problem is
described by a Luttinger liquid (LL) with an additional local
non-linear term due to backscattering from the impurity:
\beq\label{bsg}
&&S={\pi K\over 2}\int_{-\infty}^{\infty} dy\int_0^\beta
d\tau\,\left[(\partial_\tau\phi)^2+(\partial_y\phi)^2\right]\nonumber\\
&&~~~~~~~~~~~~~~~~~+2g\int_0^\beta d\tau\,\cos\left[2\pi
\phi(0,\tau)+\frac{p}{\beta}\tau\right],
\eeq
where $g$ is the amplitude of backscattering on the impurity and
$\phi(x)$ is a bosonic phase field associated with the electron
field operator $\psi(x)\sim e^{i\phi}$.  In the bosonized form the
electron-electron interaction becomes quadratic and is effectively
described by the Luttinger parameter $K$. Perturbative expansion of
the corresponding partition function in powers of $g$ produces the
series (\ref{z}) with the fugacity given by
$x=g\,\beta(2\pi/\beta\kappa)^{1/2K}$. Here $\kappa$ is a
non-universal renormalization factor, which sets the scale for the
long distance asymptotics of the correlation functions:
$\langle\exp[2i(\phi(x,\tau)-\phi(0,0))]\rangle\sim
(\kappa\sqrt{x^2+\tau^2})^{-1/K}$.  Finally, the single impurity
Kondo model is related to $Z_{p}(K,x)$ as well~\cite{FLS2}.

It is easy to understand the origin of the relation between
interference experiments and a quantum impurity problem. Moments of
fringe amplitudes are determined by high order correlation functions
computed at the same time but in different points in space.  On the
other hand, expansion of the partition function for a quantum
impurity contains correlation functions computed at the same spatial
point but at different times. Lorentz invariance of the LL ensures
that the two are the same. Note that the analogue of the finite
imaging angle $\theta$ in the interference experiments is a finite
voltage in the quantum impurity problem~\cite{bazhanov98}. This
analogy can be also understood from the interchanged roles of space
and time in the two systems.

When describing interference experiments it is convenient to define
the normalized amplitude of interference fringes $\a=A_{\tilde
p}^{2}/A_{0}^{2}$.  From eq.~(\ref{HigherMoments}) we find that
$\langle \a^{n} \rangle = Z^{(p)}_{2n}$, so by measuring
the distribution function $W_p(\a)$ experimentally, we get
direct access to the partition function~(\ref{z}). We point out that
$W_p(\a)$ can be used to compute all moments of $A_{\tilde p}^{2}$,
and therefore contains information about high order correlation
functions of the interacting Bose liquids.

Using Taylor expansion of the modified Bessel function as well as
Eq.~(\ref{z}) and the fact that $\langle \a^n\rangle = Z^p_{2n}$ we
find
\be
Z_{p}(K,x)=\int_0^\infty W_p(\a)\, I_0(2x \sqrt{\a})\, d \a.
\label{z1}
\ee
Inverting Eq.~(\ref{z1}) we can express the probability $W_p(\a)$
through the partition function $Z_{p}(K,x)$. Noting that
$I_0(ix)=J_0(x)$ and using the completeness relation for Bessel
functions, $\int_0^\infty J_0(\lambda x) J_0(\lambda y) |x|\lambda
d\lambda=\delta(|x|-|y|)$, we obtain
\be
W_p(\a)=2\int_0^\infty Z_{p}(K,i x)J_0(2x \sqrt{\a})x dx,
\label{wz}
\ee It is important that the last equation has the partition function
at imaginary value of the coupling constant. This should be understood
as analytic continuation of $Z_{p}(K,x)$.

There are several ways how one can compute $W_p(\a)$. The most
straightforward approach is to evaluate the coefficients
$Z_{2n}^{(p)}$ and thus determine all the moments of the
distribution. Explicit expressions for $Z_{2n}^{(p)}$ can be
obtained using orthogonal Jack-polynomials~\cite{FLS1}. However,
each of these coefficients is given as a series of products of
$\Gamma$-functions and their evaluation becomes extremely cumbersome
for $n\geq 3$. Another approach to finding $W_p(\a)$ is to compute
$Z_{p}(K,x)$ using the thermodynamic Bethe Ansatz for the quantum
impurity problem~\cite{FLS1}. This method works only for half
integer $K$ and requires solution of coupled integral equations.
Besides, relating $Z_{p}(K,x)$ to the distribution function of
interference amplitudes requires analytic continuation of
$Z_{p}(K,x)$ into the complex plane, $x\to ix$ (see Eq.~(\ref{wz})),
which introduces additional complications.  In this paper we will
use a different method, which is based on studies of the integrable
structure of conformal field theories~\cite{BLZ1-3}. In particular,
it was shown that the vacuum expectation value of Baxter's ${\bf Q}$
operator, central to the integrable structure of the models,
coincides with the grand partition function of interest up to an
overall prefactor~\cite{BLZ1-3} (see supplementary material for more
details):
\be
Q_p^{\rm vac}(c,\lambda)=\lambda^{4p/K} Z_{p}(K,-i x),
\label{al}
\ee
where $x$ is related to the spectral parameter $\lambda$ as
$x=\pi\lambda/ \sin(\pi/2K)$ and the central charge
$c=1-6(2K-1/(2K))^{2}$. It was conjectured in
Refs.~[\onlinecite{DT},\onlinecite{BLZ}] that the vacuum expectation
value $Q_p^{\rm vac}(\lambda)$ is proportional to the spectral
determinant of the single particle Schr\"{o}dinger equation
\begin{equation}
-\partial_{x}^{2}\Psi(x)+\left(x^{4K-2}+\frac{l(l+1)}{x^{2}}\right)\Psi(x)=E\Psi(x),
\label{schrod}
\end{equation}
where $l= 4pK -\frac{1}{2}$. So, $Q^{\rm
vac}(\lambda)=\lambda^{4p/K}  D(\rho\lambda^{2})$, where $\rho
=(4K)^{2-1/K}\Gamma^2(1-1/2K)$, $D(E)$ is the spectral determinant
defined as $D(E)=\prod_{n=1}^{\infty}(1-E/E_n)$, and $E_n$ are the
eigenvalues of (\ref{schrod}). Thus, we get
\begin{equation}
Z_{p}(K,ix)=\prod_{n=1}^{\infty}\left(1-\frac{\rho\lambda^2}{E_{n}}\right).
\label{conj}
\end{equation}
To evaluate the distribution function we solve the Schr\"odinger
Eq.(\ref{schrod}) numerically. Details of the analytical treatment
and comparison with numerics will be reported elsewhere~\cite{ADGP}.
We checked the accuracy of the numerics as well as conjecture
(\ref{conj}) by comparing coefficients $Z_{2}^{(p)}(K)$ and
$Z_{4}^{(p)}(K)$ evaluated for various $K$ using (i) the spectral
determinant (e.g. $Z_{2}^{(p)}\sim\sum 1/E_{n}$,
$Z_{4}^{(p)}\sim(\sum 1/E_{n})^{2}-\sum 1/E_{n}^{2})$ ), and (ii)
the exact expressions of Ref.~[\onlinecite{FLS1}] based on Jack
polynomials. We found perfect agreement between the two methods.

\begin{figure}[ht]
\includegraphics[width=8.5cm]{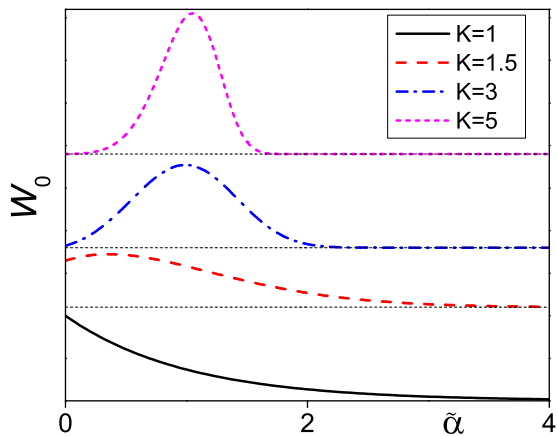}
\caption{Evolution of the distribution function $W_0(\tilde\a)$ for
different values of $K$ at $p=0$. At larger values of $K$ the
function $W_p$ tends to the delta-function (see text for
details).}
\label{figK}
\end{figure}

\begin{figure}[ht]
\includegraphics[width=8.5cm]{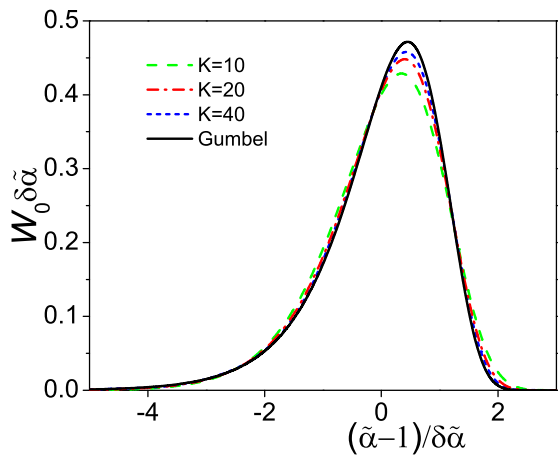}
\caption{Scaled distribution function $\delta\tilde\alpha
W_0\left((\tilde\a-1)/\delta\tilde\alpha\right)$, where
$\delta\tilde\alpha$ is the width of the distribution, for large
$K$. The function $W_0$ is multiplied by $\delta\tilde\alpha$ to
preserve the total probability, which must be equal to unity. The
dashed and dotted lines correspond to different values of $K$. The
solid line corresponds to the conjectured Gumbel distribution. }
\label{figK1}
\end{figure}

To compare distributions at different $K$ with each other it is
convenient to use a normalized interference amplitude
$\tilde\a=\a/\av{\a}=A^2/\av{A^2}$ instead $\a$. This change of
variable is also convenient for comparison with experiments. The
distribution function $W_0(\tilde\a)$ is shown in Fig.~\ref{figK}
for several values of $K$. For $K$ close to 1 (Tonks-Girardeau
limit) $W_0$ is a wide Poissonian function, which gradually narrows
as $K$ increases, finally becoming a narrow $\delta$-function at
$K\to\infty$ (the limit of noninteracting bosons). Interestingly,
the distribution function remains {\it asymmetric} for arbitrarily
large $K$. In fact we find that $W_0(\tilde\a-1)$ tends to a
universal scaling form, parameterized by a single number
characterizing the width of the distribution:
$\delta\tilde\alpha\equiv\sqrt{\langle\tilde\a^2\rangle-1}\approx
\pi/\sqrt{6}K$ (see Ref.~[\onlinecite{pad}]). We conjecture that
this limiting form of $W_0$ is the Gumbel
distribution~\cite{Katzgraber}, which frequently appears in problems
of extreme value statistics~\cite{Gumbel}: $W_0(\tilde \a-1)\to
W_G(\tilde\a,K,\gamma)$, where $\gamma\approx 0.577$ is the Euler
gamma-constant and
\be
W_G(x,a,b)=a\exp\left(a x -b - \exp(a
x-b)\right).
\ee
We plot the scaled distribution functions:
$\delta\tilde\alpha\,
W_0\left((\tilde\a-1\right)/\delta\tilde\alpha)$. Note that $W_0$
was multiplied by $\d\tilde\a$ to preserve the normalization
condition (so that the total probability is equal to unity) for
$K=10,20,40$.  For comparison in Fig.~\ref{figK1} we also present
the scaled Gumbel distribution. One can see that as $K$ increases
the function $W_0$ indeed approaches to $W_G$. Gumbel distributions
are frequently associated with random walks in strongly correlated
systems~\cite{Gumbel}. Indeed one can view the interference signal
in Eq.~(\ref{Ap2}) as a sum of contributions coming from different
points along the condensates. For weak interactions (large $K$)
these contributions are strongly correlated because the phases of
each of the condensates only weakly fluctuate along $z$. Thus there
is no surprise that $W_0$ approaches the Gumbel distribution. One
can also understand this result noting that for $K\gg 1$ the
distribution function of the interference amplitude is dominated by
rare events which reduce the contrast. The Gumbel distribution was
introduced precisely to describe rare events such as stock market
crashes or earthquakes. In supplementary materials (sec. \ref{sup1})
we also discuss distribution functions for finite values of the
observation angle (i.e. finite $p$).

Interestingly, the distribution function $W_p(\alpha)$ provides a
very simple and convenient framework for describing both the
partition function (\ref{z}) and the expectation values of the
$Q$-operator. Indeed, $W_p(\alpha)$ is a smooth well behaved
function at all values of $K$ and it can be easily approximated by
simple analytic expressions.

%


\section{Quantum simulation}

As we mentioned earlier, the distribution function $W_p(\a)$ can be
used to obtain the partition function $Z_p(K,x)$ (see Eq.~(\ref{z1}))
describing a range of various problems like quantum impurity in a
one-dimensional electron liquid, asymmetric Kondo problem, and
dissipative tunneling. It is easy to see that momentum $p$ in the
interference experiments corresponds to the external applied voltage
in the impurity problem ($ip\sim V$).  This follows from the
interchanged roles of space and time in the two problems.  Thus
measuring $W(\a)$ experimentally and taking its integral transform one
directly simulates these problems in or out of equilibrium.
Moreover, after substituting $I_0(2x\sqrt{\a})\to J_0(2x\sqrt{\a})$
in Eq.~(\ref{z1}) one obtains the partition functions of the above
models with imaginary coupling constants. Such models have been
actively investigated recently in the context of theories with
PT-symmetric rather than Hermitean Hamiltonians~\cite{bender}.

The partition function with the imaginary coupling also gives the
expectation values of the Baxter $Q$-operator (see Eq.~({\ref{al}}))
corresponding to various conformal field theories (CFTs). In
general, such theories are particularly important because many
models of two-dimensional statistical mechanics, field theory, and
many-body quantum systems at critical points can be described  by
some continuum theories having a property of conformal invariance.
This leads to a description of critical systems on the basis of
conformal field theory, which basic ingredients are the central
charge and conformal dimensions. This set of data classify different
universality classes, which often describe very different physical
models. The property of positivity of the central charge leads to a
unitary theory. Physically, the central charge determines the vacuum
(Casimir) energy of the system and governs the finite-size scaling
effects. On the other hand there is a class of models whose
universality classes of critical behavior are described by the
conformal-invariant models with {\it negative} central charges.
These theories are {\it nonunitary} and their properties are thus
very different from those described by positive central charges.

\begin{figure}[ht]
\includegraphics[width=8.5cm]{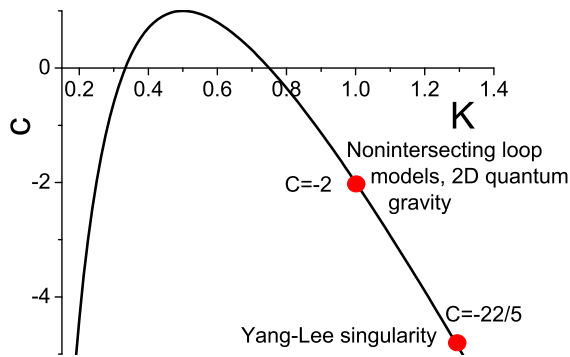}
\caption{Dependence of the central charge of conformal field
theories corresponding to interference experiments on the Luttinger
liquid parameter $K$. The red circles correspond to some particular
examples discussed in the text.}
\label{c_k}
\end{figure}

Conformal field theories corresponding to Eq.~(\ref{al}) are
characterized by {\em negative} central charge
$c=1-6(2K-1/(2K))^{2}$ and the highest weight
$\Delta=(2pK)^{2}+(c-1)/24$ (see Ref.~[\onlinecite{BLZ1-3}]). These
relations give $c\leq -2$ for $K\geq 1$. Theories with negative
central charges appear in different contexts of statistical
mechanics, stochastic growth models, 2D quantum gravity, models of
2D turbulence and even high-energy QCD. In particular, $c=-2$ CFT
has the field-theoretical representation in terms of the ghost
(anticommuting) fields and also corresponds to the critical behavior
of the non-intersecting loop model on a 2D lattice~\cite{Nienhuis}
as well as to the special case of the stochastic Loewner evolution
equation, describing the growth of random fractal stochastic
conformal-invariant interface (see e.g. Ref.~[\onlinecite{Cardy}]).
The classic example of CFT with negative $c$ is the Yang-Lee
singularity $c=-22/5$ describing the critical behavior of the Ising
model in imaginary magnetic field. Models of 2D quantum gravity
described by the fluctuating lattice geometry are related to
negative $c$ as well. Possibly, the high-energy limit of multicolor
QCD is described by the integrable CFT with negative (or zero)
central charge~\cite{Lipatov}. We illustrate the dependence of $c$
on $K$ as well as particular examples of models corresponding to
different values of $c$ in Fig.~\ref{c_k}.
The spectrum of $Q$-operator can be used to reconstruct the transfer
matrices of the above mentioned negative $c$ models. Particular
example of such procedure for $c=-2$ universality class was
explicitly constructed in Ref.~[\onlinecite{BLZ1-3}]. In this case
only the vacuum expectation value was needed to reconstruct the
whole transfer matrix. The transfer matrices contain all the
information about the properties of underlying models. In this sense
interference experiments {\it simulate} these models.
Experimentally, the central charge can be tuned by varying the
interaction $K$ and the scaling dimensions $\Delta$ of corresponding
physical operators can be manipulated by changing the observation
angle $p$. An interesting challenge here is the experimental
determination of {\it nonvacuum} values of the ${\bf Q}$ and ${\bf
T}$ operators. This is an open question.

Needless to say that the range of models mentioned above, which
belong to {\it nonunitary universality classes} is difficult (if
possible at all) to realize by other ways. The interference of
condensates provides a possible and a plausible way to explore the
interesting physics of various models ranging from statistical to
high energy physics.

\section{summary}

To summarize, in this paper we analyzed interference experiments
between two independent one dimensional quasi-condensates. We computed
the distribution function of the amplitude of interference fringes
relating this problem to the properties of $Q$ operators of conformal
field theories with negative central charges. We showed how one can
use the distribution function of the interference amplitude to reconstruct
the partition function of a two-dimensional Coulomb gas confined to a
circle. This partition function is related to a variety of
statistical and field-theoretical models. Thus studying
the interference distribution function experimentally, one can directly simulate
these interesting models.

We considered only a particular example of interference between two
one-dimensional condensates and showed the connection between the
distribution function of fringe amplitudes and the properties of
various models. This analysis can be extended to other systems with
quasi long-range order, e.g. to two-dimensional Bose systems at
finite temperature. One can expect that there will be analogous
connections to different classes of problems, some of which might
not be exactly solvable. The interference experiments open new ways
of solving these problems by direct simulation of the underlying
models.

We are grateful to I.~Affleck, C. Bender, P.~Fendley,
V.~M.~Galitski, M.~Greiner, Z.~Hadzibabic, H.~Katzgraber, M.~Lukin,
S.~L.~Lukyanov, M.~Oberthaller, M.~Oshikawa, M. Pletyukhov,
J.~Schmiedmayer, V.~Vuletic, D.~Weiss, K.~Yung and
A.~B.~Zamolodchikov for useful discussions. This work was partially
supported by the NSF grant DMR-0132874. V.G. is supported by Swiss
National Science Foundation, grant PBFR2-110423.

\section{Supplementary material}

\subsection{Mathematical details}

Function $Z(K,g)$ can be computed using the Thermodynamic Bethe
Ansatz. This was used in Refs.  [\onlinecite{FLS1,FLS2}] to evaluate
the current through a boundary impurity. However to compute the
distribution function $W$ we need analytic continuation of the
partition function $Z(K,ig)$. Expression for $Z(K,g)$ is given as a
solution of coupled integral equations and performing its analytic
continuation is not easy. In this paper we use an alternative
approach.

In a recent series of papers Bazhanov, Lukyanov, and Zamolodchikov
explored an integrable structure of conformal field theories
focusing on connections to solvable problems on lattices
\cite{BLZ1-3}. Key ingredients of solvability of lattice models are
is the so-called transfer matrix ${\bf T}(\lambda)$ operators. These
operators contain information about all  integrals of motion as well
as excitation spectra of the system. Transfer matrices are defined
as a function of the so-called spectral parameter $\lambda$ (in the
continuum limit $\lambda$ corresponds to rapidity) and commute for
different values of $\lambda$. The latter property is a direct
manifestation of the existence of infinite number of commuting
integrals of motion. In his studies of 8-vertex model,
Baxter\cite{Baxter} introduced the operator ${\bf Q}(\lambda)$ which
helps to find an eigenvalues of ${\bf T}$. Operators ${\bf T}$ and
${\bf Q}_{\pm}$ satisfy a set of commutation relations, in
particular \cite{BLZ1-3}
\begin{eqnarray}\label{BOE}
{\bf T}(\lambda){\bf Q}_{\pm}(\lambda) = {\bf
Q}_{\pm}(q\lambda)+{\bf Q}_{\pm}(q^{-1}\lambda)
\end{eqnarray}
where $q=\exp(i\pi/2K)$.  So $T$ matrices can be obtained explicitly
when one knows the Q operators.

Operators ${\bf A}_{\pm}(\lambda)={\bf Q}_{\pm}(\lambda)\lambda^{\mp
4p/K}$ act in the representation space of Virasoro algebra, which
can be constructed from the Fock space of bosonic operators $a_{\pm
n}$ satisfying $a_{n}|p\rangle=0$, $(n>0)$. The Fock vacuum state
$|p\rangle$ is an eigenstate of the momentum operator,
$P|p\rangle=p|p\rangle$. For $p=N/2$ ($N=0,1,2...$) the vacuum
eigenvalues of the operator ${\bf A}_{\pm}(\lambda)$, ${\bf
A}_{\pm}(\lambda)|p\rangle=A_{\pm}^{(vac)}|p\rangle$, are given by (
below we consider only the quantities with the $+$ subscript which
correspond to the positive $p$)
\begin{eqnarray}
A^{(vac)}(\lambda)=Z_{p}(K,-i x)
\end{eqnarray}
where
\begin{eqnarray}\label{lambda}
\lambda =\frac{x}{\pi}\sin\left(\frac{\pi}{2K}\right)
\end{eqnarray}
The function $A^{(vac)}$ has known large-$\lambda$ asymptotics
\cite{BLZ1-3}
\beq\label{asymp}
\log(A^{vac})(\lambda)\sim
M(K)\left(-\lambda^{2}\right)^{\frac{K}{2K-1}},
\eeq
where the constant $M(K)$ is given by
\beq
M(K)=\frac{\sqrt{\pi}\,\Gamma(\frac{1}{4K-2})[\Gamma(\frac{2K-1}{2K})]
^{\frac{2K}{2K-1}}}{\cos(\frac{\pi}{4K-2})\Gamma(\frac{K}{2K-1})}.
\eeq
The function $A^{(vac)}(\lambda)$ is entire function for $K>1$ and
is completely determined by its zeros $\lambda_{k}$, $k=0,1,...$.
Therefore $A^{(vac)}(\lambda)$ can be represented by the convergent
product
\begin{eqnarray}\label{entire}
A^{(vac})(\lambda)=\prod_{k=0}^{\infty}\left(1-\frac{\lambda^{2}}{\lambda_{k}^{2}}\right),
\ & &\ A^{(vac)}(0)=1
\end{eqnarray}

On the basis of analysis of a certain class of exactly solvable
model, corresponding to the integrable perturbation of the conformal
field theory, it  was conjectured in \cite{DT} that the so-called
$Y$-system and related $T$ system (where $Y=e^{\epsilon_{r}}$ and
$\epsilon_{r}$ are the Bethe-ansatz energies parametrized by $r$,
the nodes of the Dynkin diagrams) satisfy the same functional
equations and possess the same analytical structure and asymptotics
as the spectral determinant of the one-dimensional anharmonic
oscillator. Further, the same functional equations, analytical
properties (\ref{entire}) and asymptotics (\ref{asymp}) are
satisfied for the vacuum eigenvalues of ${\bf Q}$-operator for
special values of $p$ and the latter are given by the spectral
determinant of the following Schr\"{o}dinger equation
\beq
(-\partial_{x}^{2}+x^{2\alpha})\Psi(x)=E\Psi(x).
\eeq
The spectral determinant is defined as
\begin{eqnarray}\label{specdet}
D(E)=\prod_{n=1}^{\infty}\left(1-\frac{E}{E_{n}}\right).
\end{eqnarray}
Soon after, in Ref.~[\onlinecite{BLZ}], this conjecture has been
extended to all values of $p$ :
\begin{eqnarray}
A^{(vac)}(\lambda, p)= D(\rho\lambda^{2}),
\end{eqnarray}
where now $D(E)$ is the spectral determinant of Eq.~(\ref{schrod})
with $l=4pK-1/2$. Here $\rho =(4K)^{2-1/K}[\Gamma(1-1/(2K))]^{2}$.

Typically, for $n\geq 5-10$ the spectrum of the equation
(\ref{schrod}) is very well approximated by the standard WKB
expression
\beq\label{WKB}
E_{n}=\epsilon(K)(n-\gamma_{l}(K))^{\frac{2K-1}{K}}
\label{wkb1}
\eeq
where $n=1,2,..$. Here $\gamma_{l}(K)$ is the Maslov index. For
$1/2<K<\infty$, $\gamma_l(K)=\frac{1}{4}-\frac{l}{2}$, for
$K=\infty$, $\gamma_{l}(K)=-l/2$  and for $0<K<\frac{1}{2}$,
$\gamma_l(K)=\frac{4K-2l-1}{8K}$. Note that the Eq. (\ref{schrod})
has an interesting duality symmetry (generalizing the
Coulomb-harmonic oscillator duality) which allows to relate the
$K>1/2$ and $0<K<1/2$ sectors. The point $K=1/2$ is a self-dual
point of this transformation. The function $\epsilon(K)$ in
Eq.~(\ref{wkb1}) reads
\beq
\epsilon(K)=\left[\frac{2\sqrt{\pi}\,\Gamma(\frac{3}{2}+\frac{1}{4K-2})}
{\Gamma(1+\frac{1}{4K-2})}\right]^{\frac{2K-1}{K}}
\eeq
In principle, the function $\gamma_l(K)$ can be  a smooth function
interpolating between limiting values given above and can be
considered as a noninteger Maslov index~\cite{FT}. This
interpolation allows to (approximately) evaluate the partition
function and the distribution function in many cases. In the
limiting cases $K\to 1$ and $K\to \infty$ the WKB approximation
gives the exact spectrum, which was discussed in the main text. We
point however, that the using the approximate WKB spectrum can
result in non-physical results, e.g. negative values of the
distribution function $W(\a)$, and thus have to be used with care.

\subsection{Analysis of distribution functions}
\label{sup1}

\begin{figure}[h]
\includegraphics[width=8.5cm]{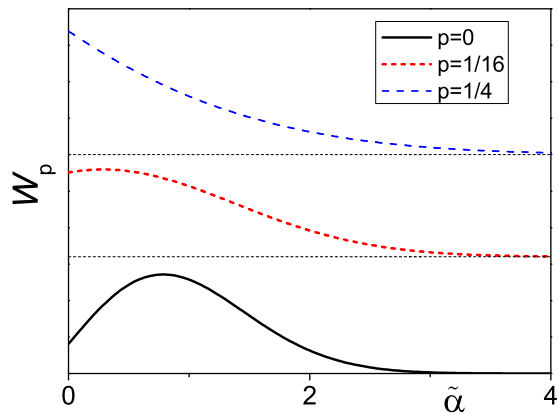}
\caption{Evolution of distribution function at $K=2$ for different
values of index $p$ (i.e. for different observation angles).}
\label{figp}
\end{figure}

Analytic expressions for $Z$ can be obtained explicitly for $K=1$
since  in this case equation (\ref{schrod}) corresponds to a
singular harmonic oscillator (with a parabolic potential replaced by
the potential $x^{2}+l(l+1)/x^{2}$, singular at $x=0$). Eigenvalues
of this Schr\"{o}dinger equation are $E_{n}=4n+2l-1$, $n=1,2,...$.
Exact result for the spectral determinant can be computed using the
Weierstrass representation of the gamma function. We find,
\begin{equation}
Z_{p}(1,ix)=\Gamma(\textstyle{\frac{1}{2}}+
2p)e^{-\frac{4C}{\pi}x^{2}}
[\Gamma(\textstyle{\frac{1}{2}}+2p-\frac{x^{2}}{\pi})]^{-1}
\end{equation}
where $C$ is a nonuniversal constant which arises due to the
logarithmic divergence of the integrals in Eq.~(\ref{z2n}) and
involves the cutoff dependence of $\langle \alpha\rangle$ at $K=1$.
If we are interested in distances much larger than the cutoff then
$C\gg 1$ and the function $Z_p$ becomes a simple gaussian. After the
integral transform this leads to the Poissonian distribution
$W_p(\a)$. The case $K\rightarrow\infty$ is recovered by the
infinite well potential for which the eigenenergies are given by
zeroes of the Bessel function. Therefore
\begin{equation}
Z_{p}(K,i x)=\Gamma(4pK+1)x^{-4pK}J_{4pK}(2 x).
\label{Kinfinity}
\end{equation}
From equations (\ref{wz}) and (\ref{Kinfinity}) one sees that when
$p=0$, the distribution function is a delta function. As $p$
increases, $W_p(\alpha)$ rapidly broadens. In the limit of large
$K$, the function $W_p$ depends only on the product $K p$ and takes a
simple form: $W_p(\alpha)\approx 4Kp\, (1-\a)^{4Kp-1}$ for $\a<1$
and zero for $\a>1$.
This function is peaked at $\a=1$ for $K p<1/4$, it becomes a step
function exactly at $K p=1/4$, and for $K p>1/4$ $W_p$ is a
monotonically decreasing function of $\alpha$. When the product $K
p$ becomes large $K p\gtrsim 1$ the function $W_p$ becomes
Poissonian: $W_p(\a)\approx 4 K p \exp(-4 K P \a)$. In general, the
tendency of broadening of the distribution function remains true for
all values of $K$. In particular, we show the behavior of $W$ at
fixed $K=2$ for various $p$ in Fig. (\ref{figp}).


\begin{thebibliography}{99}
\bibitem{ketterle}  M.~R.~Andrews, C.~G.~Townsend, H.~J.~Miesner,
D.~S.~Durfee, D.~M.~Kurn, and W.~Ketterle, Science {\bf 275}, 637
(1997).

\bibitem{pad} A.~Polkovnikov, E.~Altman, and E.~Demler,
PNAS {\bf 103}, 6125 (2006).

\bibitem{schmiedmayer}
J. Schmiedmayer et al., unpublished



\bibitem{photon} R. J. Glauber, \prl, {\bf 10}, 84 (1963).

\bibitem{meso} Ya. M. Blanter and M. B\"{u}ttiker, {\it Phys. Rep.}
{\bf 336}, 2 (2000).

\bibitem{Esslinger}
A. \"{O}ttl, S. Ritter, M. K\"{o}hl, T. Esslinger, Phys. Rev. Lett.
{\bf 95}, 090404 (2005).

\bibitem{Aspect}
M. Schellekens, R. Hoppeler, A. Perrin, J. Viana Gomes, D. Boiron,
A. Aspect, C. I. Westbrook, Science {\bf 310}, 648 (2005).

\bibitem{pra} E.~Altman, E.~Demler, and M.~D.~Lukin
Phys. Rev. A {\bf 70}, 013603 (2004).

\bibitem{greiner}  M.~Greiner, C.~A.~Regal, D.~S.~Jin, cond-mat/0502539.

\bibitem{bloch} S.~F\"olling, F.~Gerbier, A.~Widera, O.~Mandel, T.~Gericke, I.~Bloch,
Nature {\bf 434}, 481 (2005).

\bibitem{hadzibabic}
Z. Hadzibabic, P. Krüger, M. Cheneau, B. Battelier and J. Dalibard,
 preprint cond-mat/0605291, to be published in Nature.


\bibitem{levitov} L. S. Levitov, in "Quantum Noise in Mesoscopic Systems",
ed. Yu. V. Nazarov (Kluwer, 2003).

\bibitem{belzig}
W. Belzig, C. Schroll, C. Bruder, cond-mat/0412269.


\bibitem{kanefisher} C.L.~Kane and M.P.A.~Fisher, Phys. Rev. B {\bf
46}, 15233 (1992).

\bibitem{FLS2} P.~Fendley, F.~Lesage and H.~Saleur, J. Stat. Phys. {\bf 85}, 211 (1996).

\bibitem{caldeira-leggett} A.O.~Caldeira and A.J.~Leggett, Phys.
Rev. Lett. {\bf 46}, 211 (1981); Physica A {\bf 121}, 587 (1983).

\bibitem{difrancesco}
P. DiFrancesco, P. Mathieu, D. Senechal, ``Conformal
Field Theory'', Springer, New York (1999).


\bibitem{schumm}
T. Schumm, S. Hofferberth, L. M. Andersson, S. Wildermuth, S. Groth,
I. Bar-Joseph, J. Schmiedmayer, P. Krüger, Nature Physics {\bf 1},
57 (2005).


\bibitem{niu}
Q. Niu, I. Carusotto, A. B. Kuklov, Phys. Rev. A {\bf 73}, 053604
(2006).

\bibitem{cazalilla}
M. Cazalilla, J. of Phys. B: AMOP {\bf 37}, S1 (2004).

\bibitem{note} We point out that Eq.~(\ref{z2n}) corresponds to
periodic boundary contition. This may give a small quantitative
change but will not affect the qualitative picture of the evolution
of the distribution function with $K$ for long enough $L$.

\bibitem{FLS1} P.~Fendley, F.~Lesage and H.~Saleur, J. of
Stat. Phys. {\bf 79}, 799 (1995).

\bibitem{bazhanov98} V. Bazhanov, S. Lukyanov, A. Zamolodchikov,
Nucl.Phys. B {\bf 549}, 529 (1999).

\bibitem{BLZ1-3}
V. V. Bazhanov, S. L. Lukyanov, A. B. Zamolodchikov, Commun. Math.
Phys. {\bf 177}, 381 (1996); {\em ibid} {\bf 190}, 247 (1997); {\em
ibid} {\bf 200}, 297 (1999).

\bibitem{DT}
P. Dorey, R. Tateo, J. Phys. A: Math. Gen. {\bf 32}, L419 (1999).

\bibitem{BLZ}
V. V. Bazhanov, S. L. Lukyanov, A. B. Zamolodchikov, J. Stat. Phys.
{\bf 102}, 567 (2001).

\bibitem{ADGP} V.~Gritsev, E.~Altman, E.~Demler,  and A.~Polkovnikov,
in preparation.

\bibitem{Katzgraber} We thank H. G. Katzgraber for pointing out to
the Gumbel distribution.

\bibitem{Gumbel}  E.~Bertin, M.~Clusel, Journal of Physics A {\bf 39}, 7607 (2006).


\bibitem{bender} C.~M.~Bender, H.~F.~Jones, R.~J.~Rivers, Phys. Lett. B {\bf 625}, 333 (2005).

\bibitem{Nienhuis}
B. Nienhuis, Phys. Rev. Lett. {\bf 49}, 1062 (1982).

\bibitem{Cardy}
J. Cardy, Ann. Phys. {\bf 318}, 81 (2005).

\bibitem{Lipatov}
L. N. Lipatov, Phys. Rep.; L.D. Faddeev, Korchemsky, Phys. Lett. B
{\bf 342}, 311 (1995); J. Ellis, N.E. Mavromatos, Eur.Phys.J. C {\bf
8}, 91 (1999).

\bibitem{kendall} A.~Stuart, J.~K.~Ord, {\em Kendall's Advanced Theory of
Statistics}, (New York: Oxford University Press, 1987).

\bibitem{Baxter}
R. J. Baxter, Exactly Solved Models in statistical Mechanics, Acad.
Press, london (1982).

\bibitem{FT}
H. Friedrich and J. Trost, Phys. Rev. Lett. {\bf 76}, 4869 (1996).



\bibitem{affleck} I.~Affleck, W.~Hofstetter, D.~R.~Nelson, U.~Schollwock,
J. Stat. Mech. {\bf 0410}, P003 (2004).

\bibitem{note1} We note that in general even purely repulsive bosons in one
dimension can have effective positive scattering length and thus
$K<1$ (see Ref.~[\onlinecite{affleck}]).

\bibitem{DLO}
V. Dunjko, V. Lorent, M. Olshanii,  Phys. Rev. Lett. {\bf 86}, 5413
(2001).



\end{thebibliography}
\end{document}